\documentclass[review]{elsarticle}

\usepackage{amssymb}

\usepackage[english]{babel}

\journal{TBD}

\usepackage{amsmath,amssymb}
\usepackage{graphicx}\usepackage{caption}
\usepackage{color}\usepackage{bm}\usepackage{dcolumn}\usepackage{float}
\usepackage{booktabs}
\usepackage{rotating}

\usepackage[utf8]{inputenc}
\usepackage{epstopdf}
\usepackage{longtable}

\usepackage{hyperref}
\hypersetup{ colorlinks = true, citecolor = blue, urlcolor = blue}

\begin{document}

\begin{frontmatter}

\title{How Far are we from Data Mining Democratisation? \\A Systematic Review}

\author{Alfonso de la Vega, Diego Garc{\'i}a-Saiz, Marta Zorrilla, and Pablo S{\'a}nchez}

\address{Software Engineering and Real-Time Group, University of Cantabria, Santander (Spain)}

\begin{abstract}

\textbf{Context}:
Data mining techniques have demonstrated to be a powerful technique for discovering insights hidden in data from a domain.
However, these techniques demand very specialised skills.
People willing to analyse data often lack these skills, so they must rely on data scientists, which hinders \emph{data mining democratisation}.
Different approaches have appeared in the last years to address this issue.

\textbf{Objective}:
Analyse the state of the art to know how far are we from an effective data mining democratisation, what has already been accomplished, and what should be done in the upcoming years.

\textbf{Method}:
We performed a state-of-the-art review following a systematic and objective procedure, which included works both from the academia and the industry.
The reviewed works were grouped in four categories.
Each category was then evaluated in detail using a well-defined evaluation criteria to identify its strengths and weaknesses.

\textbf{Results}:
Around 700 works were initially considered, from which 43 were finally selected for a more in-depth analysis.
Only two out of the four identified categories provide effective solutions to data mining democratisation.
From these two categories, one always requires a minimum intervention of a data scientist, whereas the other one does not provide support for all the stages of the data mining process, and might exhibit accuracy problems in some contexts.

\textbf{Conclusion}:
In all analysed approaches, a data scientist is still required to perform some steps of the analysis process. 
Moreover, automated approaches that do not require data scientists for some steps expose some problems in other quality attributes, such as accuracy.
Therefore, although existent work shows some promising initial steps, we are still far from data mining democratisation.
 \end{abstract}

\begin{keyword}
Data Mining Democratisation \sep Knowledge Discovery \sep User-Centred Development \sep Self-Service Business Intelligence \sep Systematic Literature Review

\end{keyword}

\end{frontmatter}


\section{Introduction}

Currently, computer systems gather and store large amounts of data that, when properly analysed~\cite{Witten2016}, can be of great help for different purposes, which makes these data bundles highly valuable assets.
For instance, let us consider the case of \emph{Uber}\footnote{\url{https://www.uber.com/}}.
Uber is a company that offers a software system to connect particular drivers offering transport services with people who need to move around inside cities.
This system has become quite popular in some places, such as Boston or San Francisco.
As a consequence, the system has accumulated a lot of information about travel habits in the cities it operates.
This information has been found of great value to plan and improve public transport networks, and so it has been acknowledged by some city councils, which have paid Uber for access to these data~\cite{uber:bostonGlobe}.
Therefore, it can be said we are moving to a world where computers capture large amounts of data that, when properly analysed, will help take better decisions and improve systems and organizations.

Nevertheless, the analysis of gathered data is not trivial.
To use data mining techniques, a sound knowledge of their underlying mathematical and algorithmic foundations is usually required.
Unfortunately, decision makers willing to analyse a dataset, e.g., a city planner, often lack this knowledge.
Although there is a lot of research in the Knowledge Discovery and Data Mining area~\cite{Witten2016}, this community has mainly focused on finding new algorithms and methods or improving existing ones, and less attention has been paid to how these techniques can be easily used by common people, this is, people outside the data mining community~\cite{Abadi:2014,Cao2010}.
Consequently, decision makers, once they get access to a specific data source, need the help of a data scientist to process these data in order to find the information they want to obtain.
These data scientists are a scarce resource~\cite{donati2017}, which makes the analysis process slow and expensive.

\emph{Data mining democratisation} is a research area that aims to overcome this problem and make data mining and knowledge discovery techniques directly usable by people without a deep knowledge on them.
The goal of this article is to explore the state of the art of this area, by means of a systematic review, with the following goals:

\begin{enumerate}
     \item To assess whether data mining and knowledge discovery techniques are ready to be used by general decision makers.
     \item To identify strengths and weaknesses of the different approaches that constitute the current state of the art.
     \item To highlight any topics that should be addressed for the purposes of democratising data mining techniques.
\end{enumerate}

We carried out this systematic review by following the guidelines proposed by Kitchenham~\cite{Kitchenham2007}, which we complemented with the \emph{snowballing} techniques proposed by Wohlin~\cite{Wohlin2013,Wohlin2014}, so that comprehensiveness, objectivity and reproducibility of the review can be assessed.

This review complements the existing survey of \citet{Serban2013}, where the types of \emph{Intelligent Discovery Assistants (IDAs)}, these are, solutions to assist analysts in the execution of data mining processes, were described and compared.
Our work presents the following benefits over the contributions of this previous survey: (1) it updates the available information on the area, by including the latest 7 years of research (the previous survey was submitted in 2012); (2) we focus on studying the analysis democratisation issue for users without any experience in data mining, whereas most described IDAs of the previous survey are for intermediate and experienced analysts; and (3) our work was performed following and objective and systematic review method, where automated searches of scientific databases were included to offer a comprehensive view of the data mining democratisation area.

For the purposes of this review, we analysed about 700 data analysis tools and academic articles.
This combination of state-of-the-art software and research publications allows us to present a comprehensive view of what is currently being offered in terms of data mining democratisation by the industry, and of what might be available in the upcoming years from the latest works from the academia.

After this introduction, the article is organized as follows: Section~\ref{section:background} serves as background and context for this work.
Section~\ref{section:reviewmethod} describes our research questions, and the protocol that we followed when performing this systematic review.
Then, Section~\ref{section:results} comments on the obtained results, and in Section~\ref{sec:answers} we use these results to answer our initial research questions.
Finally, in Section~\ref{section:conclusions} we conclude the article by recapitulating the contributions and discoveries of this review.
 
\section{Background}
\label{section:background}

Data mining can be defined as \emph{the systematic analysis of certain data to derive useful information, which could not be directly obtained or visualized, with the objective to support decision makers work}~\cite{Witten2016}.

\begin{figure}[!tb]
  \centering
  \includegraphics[width=\linewidth]{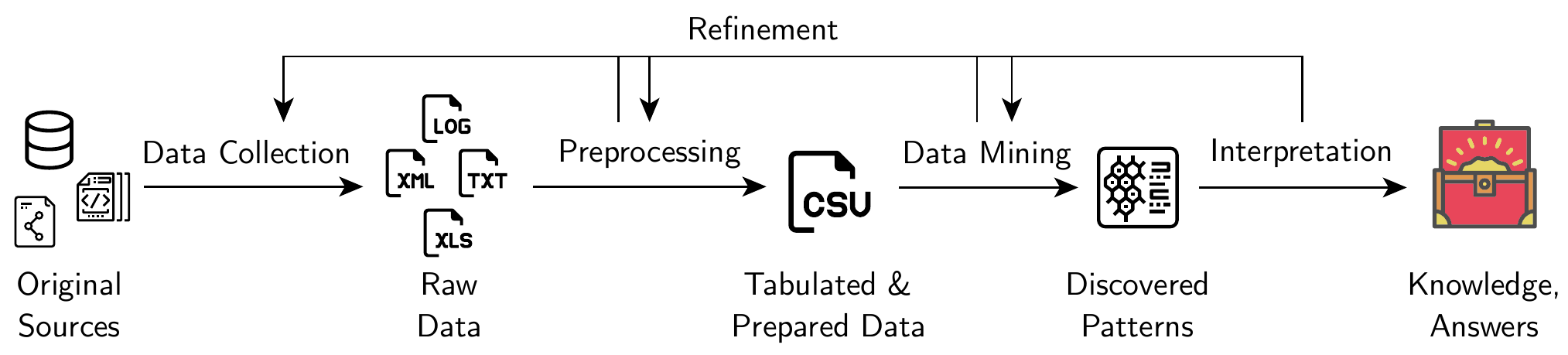}
  \caption[With footnote]{Stages of a data mining process. Source: \citet{Fayyad96fromdata}; modified for clarity. Icons designed by Smartline From Flaticon.}
  \label{fig:dataminingprocess}
\end{figure}

Data mining processes are performed to seek for answers to business questions.
Every analysis must start with a thorough comprehension about the business domain and the concrete questions to answer.
An analysis is not a monolithic process, but rather a chain of different stages applied over the data~\cite{Fayyad96fromdata}.
Figure~\ref{fig:dataminingprocess} shows a diagram of these stages, which can be summarized as follows:

\begin{description}

\item[Data Collection]
The information to analyse has to be collected from its original sources.
It may be interesting to include heterogeneous sources of information for a certain analysis, such as standard databases or real-time data streams.
Moreover, the types of the data might be different, from structured data to text documents or even media files.
As a result, the collection techniques might vary drastically depending on the input types.

\item [Preprocessing]
The obtained raw data are, in most cases, directly unusable for an analysis.
These data might contain incomplete or unstructured information, which must be cleaned and integrated into an appropriate format before being used.
In addition, it might happen that not every piece of initially collected data is relevant for an analysis.
Therefore, an exploration and selection step is required to only work with the interesting data fragments for every question to answer.
This stage usually ends with the generation of a two-dimensional (i.e. composed of rows and columns) data bundle that is ready for analysis, commonly known as a \emph{dataset}.

\item [Data Mining]
This is the stage where the proper analysis takes place.
Different algorithms are applied depending on the final goal.
For example, if we wanted to group data items according to their similarity, a clustering algorithm may be performed.
If the objective is to predict future outcomes of a variable, a classification or regression technique could be executed.
Not only it is important to select the adequate technique to employ: most data mining algorithms require the configuration of several parameters that might considerably affect its performance.

\item [Interpretation]
Once the analysis algorithms are executed, the returned results must be evaluated and interpreted.
These results may be hard to understand if they are poorly presented.
Thus, it is important to pay some attention to any generated visualizations or reports to maximize comprehensiveness.

\item [Refinement]
Frequently, when performing a step of the analysis, mistakes or previous assumptions made during the process are found; or the new insights discovered from the results of an algorithm drive the decision makers into new questions.
This is why data mining is usually defined as a cyclic and iteratively-refined process, because sometimes it is required to come back to a previous stage to fine-tune some settings before continuing with successive tasks.
\end{description}

Data mining techniques are sometimes employed in the \emph{Business Intelligence (BI)}~\cite{negash2004} area.
BI technologies are used to gain insights from data stored by a company, usually by creating reports that help decision makers visualize and understand some indicators about the performance of a business or process.
To achieve this goal, these reports aggregate data from the available sources and perform some descriptive analytics, e.g., statistics of the performance indicators.
These reports can also offer facilities to navigate through these data.
In some cases, the reports are enhanced with richer information coming from the application of data mining techniques.

A lot of data mining products exist without being noticed by final users, as they operate passively and are included in people day-to-day utilities, such as product recommendation systems in e-commerce applications.
On the other hand, there are scenarios where the user is willing to perform the mining processes actively.
Continuing with the online commerce example, a product manager may be interested in knowing the sales trend for the following term; or the profile of those clients that purchased a concrete product.

Unfortunately, most of those users who want to proactively analyse their data lack of the required knowledge to perform the data mining process described above.
Thus, there exists a gap between data mining techniques and the people who want to employ them~\cite{Cao2010}.
In the last years, several researchers have tried to democratise data mining techniques by filling this gap (e.g.~\cite{Zorrilla2013a,reif_automatic_2014,Cao2016}, among others).
Similarly, in the Business Intelligence community, several researchers and practitioners has started to work in a new area called \emph{Self-Service Business Intelligence (SSBI)}~\cite{imhoff2011,Alpar2016,Bani-Hani2018}, which aims to provide decision makers with user-friendly tools to create BI reports by themselves.

The goal of this work is to analyse the state of the art of these areas in order to know how far we are from democratising data mining, and to determine which remaining issues need to be addressed to achieve it.
Next section describes the review method that we used to perform this analysis.

\section{Review method}
\label{section:reviewmethod}

This section describes the review protocol we employed to study the current state-of-the-art of data mining democratisation.
To ensure comprehensibility, objectivity and reproducibility of this work, this protocol was designed according to the guidelines proposed by Kitchenham~\cite{Kitchenham2007} and Wohlin~\cite{Wohlin2013}.

\begin{figure}
  \centering
  \includegraphics[width=0.8\linewidth]{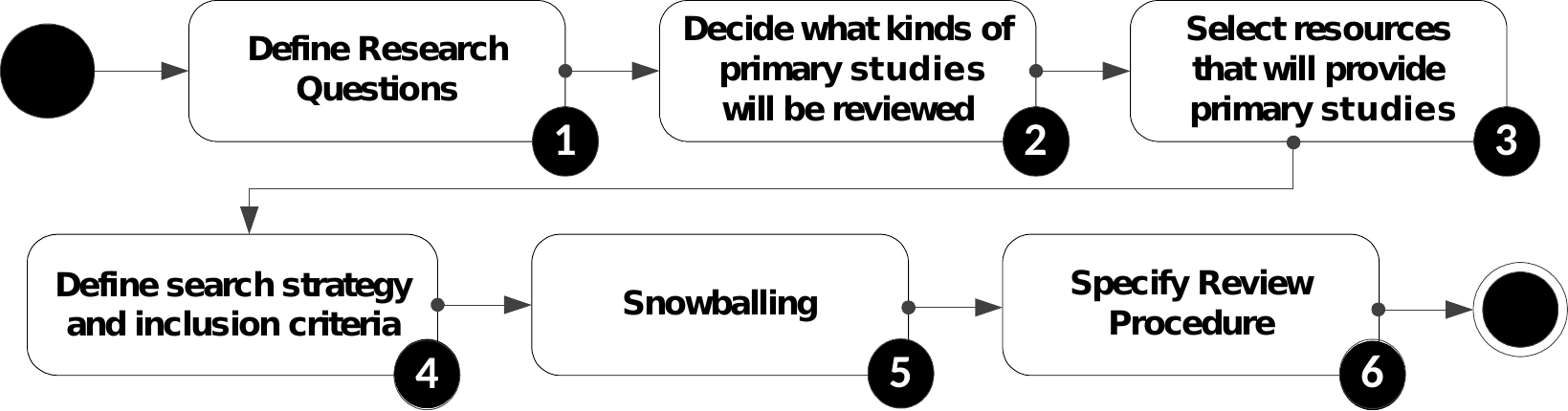}
  \caption{Process for the development of the review protocol.}
  \label{fig:reviewProcedure}
\end{figure}

The review process of this protocol is described in Figure~\ref{fig:reviewProcedure}, and it can be summarised as follows:

\begin{enumerate}
    \item First of all, the research questions that should be answered after conducting the review are defined.
    \item Then, based on these research questions, the kinds of \emph{primary studies}, e.g., research papers, that will be reviewed are determined.
    \item Next, the resources that will be used to find these primary studies, such as digital libraries, are identified.
    \item For each selected resource, a \emph{search strategy} is created.
    This search strategy must define an \emph{inclusion criteria}, which specifies precisely and objectively the reasons why a primary study should be initially considered for inclusion in the review.
    Then, each primary study is individually analysed to check whether it adheres to the purpose of this review.
    If it does not, the study is excluded.
    The reasons behind these exclusions are specified in an \emph{exclusion criteria}.
    \item Finally, to ensure comprehensiveness of our search, the selected primary studies are used as input of a \emph{snowballing} process. This process analyses backward and forward references of the primary studies, to identify new studies that might had not been found using our initial search strategy.
    These new studies are checked against the defined inclusion and exclusion criteria and, if they fulfil these criteria, they are added to the list of primary studies to be analysed.
    Then, the included studies are used as input for a new iteration of the snowballing process.
    If, after an iteration, no suitable primary studies are found, the process stops.
    \item Finally, a precise and unbiased evaluation procedure for assessing each selected primary study and answering the research questions is defined.
\end{enumerate}

Next subsections provide more details about how each one of the steps of Figure~\ref{fig:reviewProcedure} was accomplished.

\subsection{Step 1: Research Questions}
\label{sec:researchquestions}

\begin{table}
  \caption{Research questions to be answered by this systematic review.}
  \centering
  \begin{tabular}{lm{0.75\linewidth}}
  	\toprule
  	RQ0 & What approaches tackle the problem of data mining democratisation?                                                        \\
  	RQ1 & When using the approaches identified in \texttt{RQ0}, what actions do decision makers need to carry out to analyse a dataset?   \\
  	RQ2 & What technical knowledge is required to carry out the actions?                                                            \\
  	RQ3 & Can non-expert users make use of data mining tools and techniques by themselves?                                          \\
  	RQ4 & What trade-offs need to be considered for achieving data mining democratisation?                                                 \\
  	RQ5 & What should be improved in current state-of-the-art so that decision makers can properly analyse datasets by themselves? \\ \bottomrule
  \end{tabular}
  \label{table:rq}
\end{table}

Table~\ref{table:rq} shows the research questions that this systematic review aims to answer.
The ultimate objective of this review is to know how far we are from data mining democratisation (\texttt{RQ3}), and what should be done to reach that goal (\texttt{RQ5}).

To gather evidence for answering these high-level questions, we started by answering first the more fine-grained questions \texttt{RQ0-RQ2}.
\texttt{RQ0} aims to determine the size and maturity of the data mining democratisation community.
Assuming that there are approaches that tackle this problem, \texttt{RQ1} aims to identify the steps that decision makers need to accomplish to analyse a dataset by themselves.
Based on this information, \texttt{RQ2} aims to identify the minimum skills that decision makers need to have to perform an analysis by themselves.
The answers to these questions will determine whether inexperienced decision makers can be expected to properly analyse datasets without the help of a data scientist, answering \texttt{RQ3}.

To achieve data mining democratisation, some trade-offs between quality attributes need to be addressed.
For instance, data mining algorithms can be made more accessible to non-experts by preconfiguring some of their parameters.
On the other hand, this fixed pre-configuration might reduce the accuracy of the algorithms for some concrete analysis~\cite{wolpert96}.
\texttt{RQ4} explores how each approach deals with these issues.

Finally, using the answers to \texttt{RQ3} and \texttt{RQ4}, it would be interesting to identify any limitations in the current state-of-the-art that should be addressed to improve the situation of this field.
Because of this reason, \texttt{RQ5} have been added to our research questions.

Next subsection specifies the kind of materials that will be considered as \emph{primary studies} to provide an answer to these questions.

 
\subsection{Step 2: Types of Primary Studies}

To answer the previous questions, two kinds of primary studies were considered: (1) state-of-the-art-data analysis tools; and (2) research articles on data mining democratisation.

By reviewing state-of-the-art data analysis tools, we expected to get an overview of what a decision maker can currently do with these off-the-shelf software solutions; whereas the review of research literature should provide us a vision of what might analysis tools be able to do in the near future, when existent research results of the academia are transferred to the industry.

 
\subsection{Step 3: Search Resources}

We used different resources depending on the kind of primary studies that we were looking for.
The following describes these resources.

\subsubsection{Data Analysis Tools}

For finding data analysis tools, typical resources, such as scientific databases, e.g., \emph{Scopus}, were not helpful.
This was expected, since tools are rarely reported as scientific articles and, consequently, they are not contained in these databases.

Therefore, we opted for carrying out a survey among several experts in the area, to discover how to perform a systematic and comprehensive search of these tools.
Almost all of these experts recommended us to use the \emph{KDnuggets}\footnote{\url{https://www.kdnuggets.com}} website.
This website maintains highly comprehensive and up-to-date lists with more than 100 data analysis tools and libraries.
After checking the completeness of these lists, we decided to use them as the resources for finding the tools that would be reviewed.
Precisely, we used the following lists:

\begin{enumerate}
  \item The main tools list\footnote{\url{https://www.kdnuggets.com/software/suites.html}}, which contains both commercial and free/open-source software applications for data analysis.
  \item A list enumerating software that performs \emph{Automated Machine Learning}\footnote{\url{https://www.kdnuggets.com/software/automated-data-science.html}}.
  Solutions of this kind aim at automatically providing data analysis assets, e.g., prediction models, without the intervention of an expert.
\end{enumerate}

\subsubsection{Research Articles}

\begin{table}[!tb]
  \centering
  \caption{Candidate scientific databases, with their search results.}
  \begin{tabular}{lr}
    \toprule
    Database                    & \#Search results \\ \midrule
    ACM Digital Library         &              238 \\
    IEEE Xplore Digital Library &              174 \\
    INSPEC                      &              192 \\
    Science Direct              &             1357 \\
    Scopus                      &              491 \\
    Springer Link               &             5456 \\
    Web of Science              &              190 \\
    Wiley Online Library        &              324 \\ \bottomrule
  \end{tabular}
  \label{table:rawresults}
\end{table}

For the discovery of research articles, and according to the guidelines provided by Brereton~\cite{Brereton2007}, we defined a preliminary list of scientific databases for performing an automated search.
These databases are shown in Table~\ref{table:rawresults}.
Moreover, as recommended by Webster~\cite{Webster2002} and Jorgersen~\cite{Jorgensen2007}, manual search methods were used to find research works published in conferences, workshops or other venues, as some of these venues might not be indexed by scientific databases.
To find these works, a list containing the main conferences on data mining and knowledge discovery was elaborated with the collaboration of external and independent researchers of the area.
This list was complemented with some workshops specifically related to the topics of this survey.
Table~\ref{table:conferences} shows the list of selected venues for the manual search of primary studies.

\begin{table}[!tb]
    \centering
    \caption{Conferences and workshops used as resources in the manual search.}
    \begin{tabular}{lm{0.8\linewidth}}
    \toprule
    Conf 01 & European Conference on Machine Learning and Principles and Practice of Knowledge Discovery  (ECML/PKDD)\\
    Conf 02 & International Conference on Data Mining (ICDM) \\
    Conf 03 & Conference on Information and Knowledge Management (CIKM) \\
    Conf 04 & Pacific Asia Conference on Knowledge Discovery and Data Mining (PAKDD) \\
    Conf 05 & SIG Conference on Knowledge Discovery and Data Mining (SIGKDD) \\
    Work 01 & Languages for Data Mining and Machine Learning (LML) \\
                  \bottomrule
    \end{tabular}

    \label{table:conferences}
\end{table}

Summarising, three kinds of resources were used for finding the elements to be reviewed in this work: (1) the lists from the \emph{KDnuggets} website; (2) a set of scientific databases; and (3) a list of conferences and workshops.
Next section describes how our search was carried out using these resources.
 
\subsection{Step 4: Search Strategy and Selection Criteria}
\label{sec:inclusionCriteria}

Next sections describe how each resource was individually processed, according to its own particularities, to find those studies that were later reviewed.

\subsubsection{Data Analysis Tools}

Each data analysis tool from the lists provided by \emph{KDnuggets} was initially considered as a potential primary study for the review.
Thus, the \emph{inclusion criteria} for these tools was simply their appearance in the selected lists, which gave us a total of 138 candidates to check.
This number corresponds with the last time we audited the lists (June 2018).

\begin{table}
  \centering
  \caption{Exclusion criteria for data analysis tools.}
  \begin{tabular}{ll}
    \toprule
    EC1.1 & The tool is deprecated. \\
    EC1.2 & The tool requires advanced computing skills to be used. \\
    EC1.3 & The tool requires some customer-specific development. \\
    \bottomrule
  \end{tabular}
  \label{table:toolsExlusionCriteria}
\end{table}

The tools were reviewed individually, to discard those that were not helpful for the purpose of this review.
Tools were discarded when they exhibited one or more items of the exclusion criteria depicted in Table~\ref{table:toolsExlusionCriteria}.

Deprecated tools that are not longer maintained were discarded (\emph{EC1.1}), since we understood that this deprecation was either because the tool was not useful at all, or because it had been superseded by a similar tool.
Therefore, the analysis of this posterior, more successful tool should be enough.

In addition, tools that require advanced computing skills were also discarded (\emph{EC1.2}).
For instance, programming libraries for knowledge discovery such as MLC++~\cite{Kohavi1996} were removed.
These tools are designed specifically for developers and programmers, and not for being used by decision makers, so they are out of the scope of this review.

Finally, it was detected that some tools from the \emph{KDnuggets'} lists were not tools exactly, but companies that offer some services.
As an example, \emph{ThinkAnalytics} is a company specialized in recommender systems.
If a business user wants to acquire its product, they must contact this company, which will customize it for them.
However, the product cannot be acquired without the customisations.
These customisations would be similar to the process of hiring a data scientist to analyse a dataset on behalf of a decision maker, that is what we try to avoid in this review.
Therefore, this kind of tools was also discarded (\emph{EC1.3}).

A list of the reviewed tools and the reasons why they were added or excluded from the review can be found as supplementary material of this review.
After this step, 28 tools were selected as primary studies for further analysis.

\subsubsection{Scientific Databases}
\label{sec:selectionCriteriaAcademia}

\begin{table}
  \centering
  \caption{Search string used in the scientific databases.}
  \begin{tabular}{l l}
  	\hline
  	Major Terms           & Search Terms                                       \\ \hline
  	Data Mining           & \texttt{(("data mining" OR "knowledge discovery")} \\
  	                      & \texttt{~AND}                                       \\
  	Usable by non-experts & \texttt{~("democrati*" OR "non-expert*" OR}         \\
  	                      & \texttt{~"user oriented" OR "user-oriented" OR}     \\
  	                      & \texttt{~"user centered" OR "user-centered"))}      \\
  	                      & \texttt{OR}                                        \\
  	Business Intelligence & \texttt{"self-service business intelligence")}     \\ \hline
  \end{tabular}
    \label{table:searchString}
 \end{table}

According the guidelines provided by Kitchenham~\cite{Kitchenham2007} and Wohlin~\cite{Wohlin2012}, a search string for executing an automated search in the scientific databases was constructed.
This string is depicted in Table~\ref{table:searchString}.
To avoid bias and ensure comprehensiveness, this search string was submitted for review and approval to two external data mining experts.

The goal of this search string was to retrieve articles where: (1) either the final users were taken into account when developing a data analysis system; or (2) these final users were able to tweak some aspects of the data analysis process by themselves.
This search string was iteratively constructed and refined.
First, as many related terms as possible were included to make the search highly comprehensive.
However, the number of returned results was extremely high, and these results included a lot of work that was not related to the topic of this review.
For instance, the inclusion of terms describing easiness of use, such as ``friendly'', ``user-friendly'' or ``usable'', added a cumbersome number of articles which were outside the scope of this review.
So, these terms were skipped to increase accuracy of the results.

We applied the search string in the selected databases to the title, abstract, and keywords of scientific articles.
For the sake of comprehensiveness, the search was not limited to any particular discipline, as recommended by Kitchenham~\cite{Kitchenham2007}.
For instance, articles related to the topic of this review might be published in medical journals.
Moreover, the search was limited to those articles that, in addition, satisfied the following inclusion criteria: we included peer-reviewed articles, written in English, whose publication date happened up to June 2018.

Table~\ref{table:rawresults} shows, besides each considered database, the number of results returned for our search string.
As it can be observed, some databases, such as \emph{Science Direct} or \emph{Springer Link}, returned a very large number of results.
Nevertheless, most of these were not of interest for our review.
For instance, the results included topics such as mutators for genetic programming or latency-based issues of wireless network, which are not connected to the topic of this review.
So, we opted for using a subset of these candidate databases, which offered a good balance between accuracy of results and coverage of scientific journals and conference proceedings.

With this premise in mind, \emph{Scopus}, \emph{Web of Science} and \emph{INSPEC} were selected.
Before discarding \emph{Science Direct}, \emph{Wiley Online Library}, \emph{ACM Digital Library}, \emph{IEEE Xplore Digital Library}, and \emph{Springer Link}, it was checked that relevant journals and conference proceedings indexed by these databases were also indexed by the ones we selected.

The selected databases returned an initial number of 873 articles.
After a cleaning process, where we removed duplicated articles and most invalid results (e.g., table of contents of some conferences showed up as result entries), 559 articles were finally selected as candidate primary studies.
These articles are listed in the complementary material.

\begin{table}[!tb]
  \centering
  \caption{Exclusion criteria for articles in scientific databases.}
  \begin{tabular}{ll}
        \toprule
        EC2.1 & The work is a position paper. \\
        EC2.2 & The work does not address any steps of a data mining process. \\
        EC2.3 & The work is not oriented to users outside the data mining area. \\
        EC2.4 & The work is not designed to be used with arbitrary datasets.  \\
        \bottomrule
    \end{tabular}
    \label{table:databasesExclusionCriteria}
\end{table}

The candidates were individually reviewed to select those that fitted with the purpose of the review.
The selection process can be summarised as follows:

\begin{enumerate}
    \item First, we read the abstract of each article.
    Those articles that were considered clearly out of scope were discarded. When in doubt, articles were included for further analysis.
    \item Then, for the remaining articles, we obtained and read the full versions of each work.
    Again, those articles that did not fit with the purpose of the review were eliminated.
    \item Finally, articles written by the same authors and featuring the same line of research were grouped, and the most mature and comprehensive work of each group was selected.
\end{enumerate}

Table~\ref{table:databasesExclusionCriteria} specifies the exclusion criteria that was used for discarding research works.
First of all, position papers that just state the need for data mining democratisation but that do not describe any approach to achieve it, e.g., ~\cite{Lu2017,Vanwinckelen2013}, were left out of the review.

Secondly, we were interested in works about data mining.
We did not require the contributions of the selected works to address the whole data mining process, but at least they must address one step of this process, such as data preprocessing, or algorithms selection and execution.

The third criteria for exclusion is determined by the review's focus: we discarded those works that showed clear indicators of being oriented for experts, e.g. articles describing the internals of new analysis algorithms, or presenting utilities that required the knowledge of advanced data mining concepts for their configuration and usage.

Finally, our preliminary searches detected some articles that described software applications for the analysis of data from a concrete domain.
These applications were designed to be used for experts in that domain, who had no knowledge in data mining.
Therefore, a special effort was made to hide any low-level analysis details to these users.
Data mining experts were the ones developing these applications.
In the development, these experts addressed exclusively a very specific problem of a concrete domain, without aiming to make the resulting application reusable for other domains or datasets.
For this reason, we considered that these approaches, which we will refer to in the following as \emph{ad-hoc applications}, do not fit at all with the purpose of this review, and they cannot be properly analysed using the review procedure that we describe in the next section.
Therefore, these approaches were discarded.
Nevertheless, there are some concrete contributions of these works that might help achieve data mining democratisation.
To make this review more comprehensive, and in case the reader is interested, these contributions are summarised in \ref{sec:domainspecificapps}.

After the end of the automated search, 15 articles that address the data mining democratisation issue from a broad and generic perspective were accepted as primary studies for this review.
In addition, we detected 11 ad-hoc applications that fall under the description of the previous paragraph.

\subsubsection{Conferences and Workshops}

Before looking at the proceedings of the conferences and the workshop listed in Table~\ref{table:conferences}, we checked whether these proceedings were already indexed in the \emph{Scopus}, \emph{Web of Science} or \emph{INSPEC} databases, to avoid doing redundant work.
All of them were already indexed by these databases, but the proceedings of the 2015 edition of ICDM (\texttt{Conf 02}) and the workshop proceedings. So, we reviewed these venues manually.
Each article was checked against the exclusion criteria contained in Table~\ref{table:databasesExclusionCriteria}.
As a result, one article was selected for full review~\cite{Vanwinckelen2013}, but it was later excluded because it was oriented for computer and statistics-savvy users.

 
\subsection{Step 5: Snowballing}

To ensure comprehensiveness of our search process, and following the guidelines proposed by Jalali et al.~\cite{Jalali2012}, we used the initially selected 26 primary studies from the automated search as input of a \emph{snowballing process}.
During this process, articles cited by and citing the primary studies were analysed.
The goal of this process was twofold:

\begin{enumerate}
    \item To discover any work that should be included in the review, but that had not been retrieved by the automated search.
    \item To find follow-up articles of primary studies reporting a more mature work.
\end{enumerate}

The process was achieved by checking one level of backward and forward  references of the primary sources. \emph{Scopus} and \emph{Google Scholar} were used to find the forward references, i.e., articles citing the primary studies. For each backward or forward reference, the exclusion criteria of Table~\ref{table:databasesExclusionCriteria} was applied.

As a result of the first iteration of this process, two new articles~\cite{Chittaro2003,Guyet2007} were added to the list of ad-hoc applications.
In addition, an article describing another ad-hoc application~\cite{peng_user-oriented_2014} was superseded by a more recent publication~\cite{Peng2015} of the same authors.

The snowballing process was repeated using the newly found articles as input.
As a result of this second iteration, no new work was identified, which implied the stop of the snowballing process, and the end of our article search.
Finally, a total of 28 articles was found in our search, of which 15 articles were selected as primary studies for their review, and 13 articles conformed the list of ad-hoc applications (see \ref{sec:domainspecificapps}).

The selected 15 articles, combined with the 28 tools previously selected from the \emph{KDnuggets} lists, add up to 43 primary studies that were analysed in this work.
Next section describes how this analysis was carried out.

\subsection{Step 6: Evaluation Procedure}
\label{sec:evaluationProcedure}

A systematic and objective procedure was designed to analyse the selected primary studies.
This procedure consisted on gathering a set of indicators, which were grouped into two main categories.
These categories are described below. 
\subsubsection{Assistance During the Data Mining Process}

The objective of this first set of indicators was to provide an answer for the research questions \texttt{RQ1} and \texttt{RQ2}.
For each approach, and for each stage of a data mining process (see Section~\ref{section:background}), questions of Table~\ref{table:eqAssistance} were answered.
Refinement of results, this is, the possibility of exploring new issues based on results of a data mining process, was also considered as another stage of the data mining process.

\begin{table}
  \centering
  \caption{Questions to assess stage assistance during a data mining process.}
  \begin{tabular}{l p{0.8\linewidth}}
      \toprule
      EQ1.1 & Is this stage covered by the approach? \\
      EQ1.2 & How is the decision maker assisted during this stage? \\
      \bottomrule
  \end{tabular}
  \label{table:eqAssistance}
\end{table}

\subsubsection{Analysis of Trade-Offs}

As commented in Section~\ref{sec:researchquestions}, achieving data mining democratisation implies facing trade-offs between different quality attributes.
So, we analysed how a set of these attributes are satisfied by each selected primary study.
More specifically, we focused on those quality attributes that typically conflict in the case of data mining democratisation: adoption cost, accuracy of the solutions, functional completeness (i.e. what types of data mining techniques are offered by each approach, and what types are not), and evolution capabilities.
These quality attributes, as well as the questions we performed during the data collection, are described in Table~\ref{table:tradeoffs}.
The results of this analysis should provide an answer to the research question \texttt{RQ4}.

\begin{table}
    \centering
    \caption{Evaluation questions for the quality attributes analysis.}
    \begin{tabular}{l l p{0.6\linewidth}}
        \toprule
        Quality Attribute & \multicolumn{2}{l}{Questions} \\
        \midrule
        Adoption Cost & EQ2.1 & Can the approach be deployed in a new domain without requiring some adaptations? \\
        & EQ2.2 & Can non-experts carry out any required adaptations without the help of an expert? \\
        \midrule
        Accuracy & EQ2.3 & Are the provided results as accurate as possible, i.e., can they be similar to what an expert could achieve manually? \\
          & EQ2.4 & Can the analysis be tuned to produce more accurate or precise results? \\
          & EQ2.5 & Can non-experts perform that tuning by themselves? \\
        \midrule
        Completeness & EQ2.6 & What analysis techniques are available in the approach?\\
        \midrule
        Evolvability & EQ2.7 & How easy it is to extend the approach to cover new user needs? \\
          & EQ2.8 & Can new analysis techniques be incorporated into this approach? \\
        \bottomrule
    \end{tabular}

    \label{table:tradeoffs}
\end{table}

\section{Results}
\label{section:results}

This section describes the results gathered after reviewing and evaluating the selected primary studies.
For this purpose, we distributed these studies into four categories, which were then analysed.
First, we show a summary of these categories, and give some explanation on how primary studies were grouped.
Then, each category is evaluated using the procedure described in Section~\ref{sec:evaluationProcedure}.

\subsection{Classification of Selected Studies}
\label{sec:resultssummary}

For the sake of brevity, we do not describe each selected primary study individually.
Instead, these studies were grouped according to their similarities, and then they were analysed and compared as groups.
We identified four different groups or categories, which are described in the following sections.
The correspondence of each primary study with its category is provided in Table~\ref{table:categories}.

\begin{table}
	\centering
	\caption{Categories of the approaches that address data mining democratisation.}
    \begin{tabular}{m{0.2\linewidth}m{0.7\linewidth}}
    \toprule
    Category & Primary studies \\
    \midrule
        Workflow-based tools & AdvancedMiner,
                           Alteryx,
                           Angoss Knowledge Studio,
                           BDB Predictive Workbench,
                           Coheris SPAD,
                           Dataiku,
                           Exeura Rialto,
                           IBM SPSS,
                           KNIME,
                           Orange,
                           Partek,
                           Rapidminer,
                           Weka.
                           \\
        Self-Service Business Intelligence &  \citet{Schuff2018},
                                          \citet{Behringer2017},
                                          \citet{Sulaiman2016},
                                          \citet{Schlesinger201511},
                                          \citet{Abello2013},
                                          Microsoft Power BI,
                                          Tableau,
                                          IBM Watson Analytics.
                                          \\
        Black-Box Components  &  \citet{campos_data-centric_2005},
                             \citet{reif_automatic_2014},
                             \citet{ankerst_towards_2000},
                             \citet{Bilalli2017},
                             \citet{Han2015a},
                                                          Automatic Business Modeler,
                             AutoDiscovery,
                             Auto-Weka,
                             Bicedeep AI,
                             DataRobot,
                             DMWay,
                             Emcien,
                             ForecastThis DSX,
                             Featuretools,
                             Kogentix,
                             MLJAR,
                             Xpanse Analytics.
                             \\
        Development \hspace{2cm} Frameworks
                           &  \citet{ben_ayed_user-centered_2010},
                              \citet{espinosa_enabling_2015},
                              \citet{Zorrilla2013a},
                              \citet{Alonso2018},
                              \citet{santos_data_2013}.
                              \\
    \bottomrule
    \end{tabular}
    \label{table:categories}
\end{table}

\subsubsection{Workflow-Based Tools}

This category is composed entirely of state-of-the-art tools that are based on the following approach: they provide a set of building blocks, where each block performs an analysis-related task.
Users connect these blocks graphically to create what is known as a \emph{workflow}, which specifies a data mining process.
The idea is that data mining processes can be specified faster and in a more friendly way by means of dragging and dropping some of these prebuilt blocks, which are later tuned according to the particularities of each concrete analysis.
To support this tuning, building blocks have some configurable parameters that can be adjusted.

\begin{figure}
  \centering
    \includegraphics[width=0.85\linewidth]{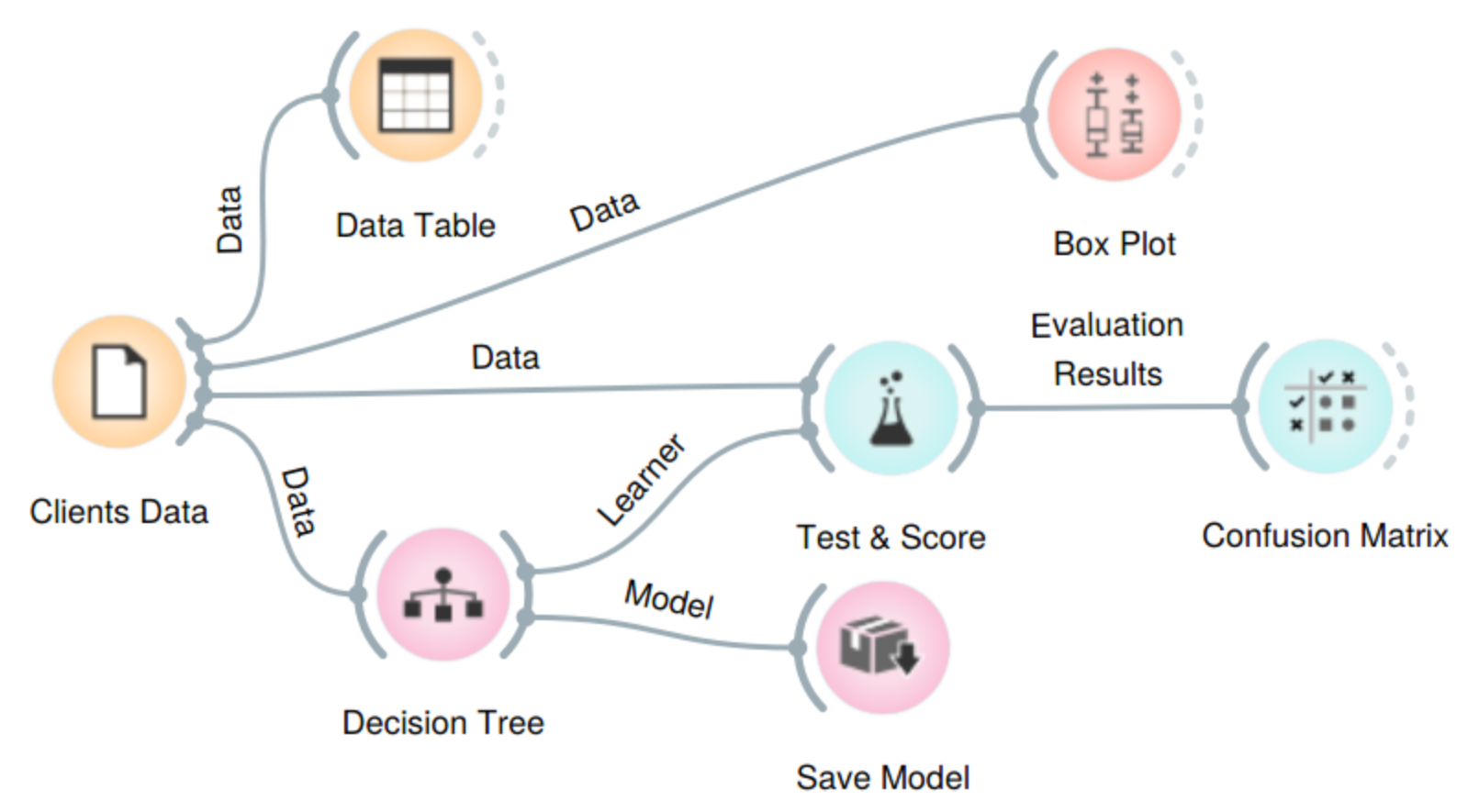}
  \caption{Data mining process example specified with an Orange workflow.}
  \label{fig:orangeExample}
\end{figure}

To illustrate this category, we have selected \emph{Orange}\footnote{\url{http://orange.biolab.si}} because, in our humble opinion, it is one of the most usable tools and a well-known representative of this category, and it is freely available.
Figure~\ref{fig:orangeExample} shows an Orange workflow example specifying a data mining process.

In this workflow, information about clients of a bank is used to predict whether they will subscribe to a term deposit offer before contacting them via phone call.
A dataset containing historical information about the clients (e.g. age, current balance, has an mortgage, credit status, whether previous marketing calls were successful) is loaded with a \emph{File} block, which in the workflow of the figure has been renamed to \emph{Clients Data}.
The contents of this dataset can be visualised using a \emph{Data Table} block.
More detailed information about these data can be obtained with the \emph{Box Plot} block that, as it could be expected, draws box plots of each column.
Based on the loaded dataset, the \emph{Decision Tree} learner block trains a prediction model, which can be used to estimate the answer of a client to the subscription suggestion.
To evaluate the prediction accuracy of this model, the \emph{Test \& Score} block performs a cross-validation process (see Chapter 5.3 of~\cite{Witten2016}) with the provided data.
The \emph{Confusion Matrix} shows the comparison of the obtained predictions against the correct ones in a compact and friendly way.
When satisfied with the performance of the learner, it can be stored for future usage with the \emph{Save Model} block.

Each block offers some configurable parameters to tune the analysis process.
For instance, for the \emph{Decision Tree} block, a maximum depth of the output tree can be specified in order to avoid creating too large and overfitted decision trees.
Similarly, for the evaluation block, the number of folds used by the cross-validation process might be changed.

Just from the concepts appearing in the description of the previous example, it can be perceived than a sound knowledge of the techniques employed by these blocks might be required to appropriately select, connect and configure them.

\subsubsection{Self-Service Business Intelligence}
\label{sec:selfservicebusinessintelligence}

In Section~\ref{section:background}, we introduced the term \emph{Self-Service Business Intelligence (SSBI)} \cite{imhoff2011,Alpar2016,Bani-Hani2018}, which
refers to a research and industry field where decision makers that use BI services are provided with user-friendly tools to create reports by themselves, making unnecessary the intervention of BI experts. Several companies are adapting their BI solutions to include self-service components for non-expert users.
Representatives of these solutions, such as \emph{Tableau}, have been studied and included in Table~\ref{table:categories}.

Currently, SSBI tools allow decision makers to perform the following process:
First, data is loaded into the tool through the appropriate data source connectors.
These connectors allow extracting data from different sources, such as a relational database or an Excel sheet.
Then, the imported raw data is filtered and processed.
Finally, the information is organized into a report or dashboard where it can be easily visualised and digested.

Continuing with the example of bank marketing calls, using BI, we could load a dataset containing the data of the clients with a CSV (\emph{Comma-Separated Values}) file connector.
Then, we might filter the data to focus only on clients that previously accepted a term deposit offer.
A report using textual explanations combined with graphs like \emph{scatter plots} or \emph{dice charts} might be created to visualise key performance indicators (KPIs) such as the \emph{Cold Calling Success Ratio}, i.e., the ratio of successful calls to potential clients; aggregated marketing results by country or state; or trend estimations for future seasons.
All this process would be performed by non-experts navigating through menus and wizards in a high-level graphical interface.
This process is hard to synthesize in a single image, so a picture illustrating it such as Figure~\ref{fig:orangeExample} is not provided for the sake of simplicity.

Researchers are trying to improve the current state-of-the-art of these tools by improve the way in which end users interact with these systems.
For instance, some authors are working on making the management of the available data easier~\cite{Schlesinger201511,Abello2013,Sulaiman2016}; and on increasing the variety of analysis that these solutions can perform~\cite{Behringer2017, Schuff2018}.

In summary, this group is somehow similar to the \emph{Workflow-Based Tools}, but it is more oriented to the creation of dashboards and reports, instead of particular data mining processes.

\subsubsection{Black-Box Components}

This category groups research work and tools whose goal is to hide details of data mining processes, so that they can be carried out by non-expert users.
We named these approaches as \emph{black-box components}, because (1) they can be applied as are to new contexts, without any parameter tuning or adjustments (i.e. they offer black-box functionality); and (2) they offer support of some stages of the analysis, but not for the process as a whole, so they can participate as components of an analysis.

Most research work in this category focus on facilitating the mining stage~\cite{campos_data-centric_2005,reif_automatic_2014,ankerst_towards_2000,Han2015a}, but there are also some approaches that deal with, for instance, the data preprocessing stage~\cite{Bilalli2017}.

A clear representative of this category is~\cite{campos_data-centric_2005}, where a data mining module extension for the Oracle Database Management System is presented.
This extension offers some procedures that execute prebuilt data mining processes using as input the data contained in a Oracle relational database table.

For instance, the \emph{PROFILE} procedure allows obtaining similarities of those records which share the same value for a specific column.
Continuing with the bank marketing example, the \emph{PROFILE(clients\_information, call\_outcome, results)} command might be executed to analyse the \emph{clients\_information} table, with the objective of checking what features are shared between those clients who have the same value in the \emph{call\_outcome} column.
The results of this procedure, which are provided in the output variable \emph{results}, are two sets of rules. The first set describes those clients who are likely to accept the term deposit and the second one those ones who are not.

The main contribution of this approach is that the user does not need to know the details behind these procedures.
The user knows neither what specific algorithm is being used to execute the procedure, nor how the algorithm parameters have been exactly configured.
Nevertheless, data must be retrieved from its sources, cleaned, and processed to fit into an adequate format, which in this case is a single relational table, before these procedures can be executed.

Regarding other approaches in this group, Reif et al.~\cite{reif_automatic_2014} offers a building block which can be used in a RapidMiner workflow to automatize the selection of a classifier.
Ankerst et al.~\cite{ankerst_towards_2000} includes computer-based visualization techniques that aid the end user in the creation of a decision tree.
Han and Leung~\cite{Han2015a} presents a web service to automatically find frequent sets of items that often appears together.
Finally, Bilalli et al.~\cite{Bilalli2017} present a black-box component that, based on the characteristics of a given dataset, applies a set of preprocessing transformations to prepare the data and improve the prediction results of the generated models.
As with the previous example, the main contribution of these approaches is that some tasks of a data mining process are automated and carried out by a computer transparently.

Selected tools belonging to this category also try to automate the mining stage of the process, focusing almost completely on prediction analysis.
Users of these tools have to provide a dataset with training records, and specify what they want to predict.
Then, the tools automatically build a prediction model without the intervention of any expert.
This prediction model can then be used over new records to perform predictions.
Examples of these tools are \emph{DataRobot}, \emph{Kogentix} or \emph{MLJAR}.
As an exception, one of the tools, \emph{Featuretools}, focuses on the preprocessing stage by automating \emph{feature engineering}, i.e. the generation and selection of features that will be used to train a model.

\subsubsection{Development Frameworks}

This category contains research works that provide methodologies to develop knowledge discovery systems, with the objective of making these systems usable by non-experts~\cite{ben_ayed_user-centered_2010,espinosa_enabling_2015,santos_data_2013, Zorrilla2013a,Alonso2018}.

A representative of this category is the work of Zorrilla and Garc{\'i}a-Saiz~\cite{Zorrilla2013a}, who propose to use a service-oriented methodology to develop data analysis systems.
The first step in this methodology is to know which questions end users want to answer.
Then, the data mining processes that will compute these answers are developed by data scientists and wrapped as web services.
These wrapped processes are designed to be as automated as possible, preventing the parameters of their algorithms from needing adjustments due to slight changes in the input data.
Finally, user-friendly, web-based interfaces are developed.
These interfaces must allow non-experts to invoke the analysis services and receive the obtained results, without requiring any technical knowledge of the wrapped processes.

Using this approach in our marketing example, a preconfigured prediction process to analyse clients data would be initially developed by experts, and encapsulated in a web service.
Then, a user-friendly web interface to invoke this service would be developed and deployed.
Using this interface, bank employees could determine the likelihood of a client accepting their marketing offer.
Since the wrapped prediction process is autoconfigurable, it should also work with datasets coming from other banks, although data in these datasets can exhibit other internal properties, such as a higher number of outliers.

As for the other approaches of this category, Ben Ayed et al.~\cite{ben_ayed_user-centered_2010} present a development process that combines the Unified Process from Software Engineering~\cite{Kruchten2003} with the U model for Human-Computer Interaction~\cite{Lepreux2003} to generate an iterative, user-centred development method, which outputs knowledge discovery applications that can be used by non-experts.
Espinosa et al.~\cite{espinosa_enabling_2015} propose a methodology that guide users in the development of data mining applications.
This methodology is based on two main elements: (1) a taxonomy of questions that helps non-expert users to identify the data mining technique they should use, i.e., clustering to group data; and (2) a recommender system that returns the best data mining algorithm for a given technique inside a specific context (e.g. Kmeans or Expectation Maximization as clustering algorithms).

Santos et al.~\cite{santos_data_2013} present a reference architecture to build data mining systems.
This architecture includes: (1) a data warehouse that contains the data to be analysed; (2) an ontology that models the domain to which these data belongs to; and (3) a set of metadata that specifies how the ontology connects to the data warehouse and how to process each data bundle.
These metadata are specified with the help of a data mining expert.
Once these elements have been created, end users interact with the domain ontology to select those data they want to analyse.
Based on the metadata, the system proposes several kinds of analysis, and the end user selects one of them.
Then, using the metadata again, the system automatically instantiates a data mining process, configures it appropriately and executes it, returning the results to the end user.
Finally, Alonso et al.~\cite{Alonso2018} introduce a development process, based on fuzzy logic, to create data analysis systems whose results can be easily interpreted by end users, providing explanations in natural language. 

It should be noted that, although the systems generated using these approaches are initially prepared by data mining experts, once they are ready for use, these experts are not required any more.
It could be argued that these systems are similar to the \emph{ad-hoc applications} commented in Section~\ref{sec:inclusionCriteria} and \ref{sec:domainspecificapps}, i.e., applications developed by experts to solve a specific problem in a particular setting; and, therefore, they should be discarded according to our \emph{exclusion criteria} (see Table \ref{table:databasesExclusionCriteria}).
Nevertheless, oppositely to \emph{ad-hoc applications}, these approaches are designed to be applicable to different domains.
In addition, the generated applications could also be reused without further adjustments with different datasets from the same domain, whereas \emph{ad-hoc applications} are designed to work with a single and very specific dataset, without reuse across a domain (or in any other different domain) in mind.
 
\subsection{Evaluation Results}

Now we comment on the results obtained after executing the evaluation procedure described in Section~\ref{sec:evaluationProcedure} to the selected primary studies.
This evaluation procedure was divided into two stages, being each stage described in its corresponding subsection.

\subsubsection{Assistance During the Data Mining Process}

This first stage of the evaluation procedure (see Section~\ref{sec:evaluationProcedure} and Table~\ref{table:eqAssistance}) aims to know what stages of the data mining process are covered by each approach, and how well each stage is covered.
Results of this analysis are summarised in Tables~\ref{table:apaStages} and~\ref{table:apaWhatAssistance}.
Table~\ref{table:apaStages} shows the coverage of the stages of a data mining process, whereas Table~\ref{table:apaWhatAssistance} describes briefly how these stages are covered.
In Table~\ref{table:apaStages}, a stage is marked if at least one of the approaches of the corresponding category addresses it.
A check mark surrounded with parentheses indicates that limited support for that stage is present, but it is not as clear as in the other cases.
We comment on these results in the following.

\paragraph{EQ1.1: Covered Data Mining Process Stages}

\begin{table}
  \centering
  \caption{Coverage of the data mining process stages offered by each category.}
  \begin{tabular}{m{0.30\linewidth}ccccc}
  	\toprule
  	Category$\backslash$Stage Action & Collect  & Preprocess &   Mine   & Interpret & Refine \\ \midrule
  	Workflows                 & \checkmark &  \checkmark   & \checkmark &   \checkmark   & \checkmark \\
  	SSBI                      & \checkmark &  \checkmark   & (\checkmark) &   \checkmark   & \checkmark \\
  	Black-Box Components              &            &  \checkmark   & \checkmark &                &            \\
  	Development \hspace{2cm} Frameworks    & \checkmark &  \checkmark   & \checkmark &   \checkmark   &      (\checkmark)      \\ \bottomrule
  \end{tabular}
  \label{table:apaStages}
\end{table}

Workflow-based tools and SSBI solutions cover the whole data mining process, since they offer building blocks or wizards for all the stages.
Figure~\ref{fig:orangeExample} provides an example of a workflow where all data mining stages are covered.
Moreover, these blocks and wizards typically contain configurable parameters, so the process can be tuned and refined.

Black-box components assist either by automatically selecting data mining algorithms, or by providing users with the ability to execute analysis tasks without knowing their low-level details.
In these approaches, only the mining stage is typically covered at the time of writing this survey.
Therefore, the user is still in charge of acquiring and preparing the data, interpreting the results of the analysis, and adjusting the input data if some refinements are desired.
Exceptions are the work of Bilalli et al.~\cite{Bilalli2017} and Featuretools~\cite{kanter2015}, which try to automate the preprocessing stage.

The development frameworks require the intervention of a data mining expert to create some initial elements.
Nevertheless, once these elements are ready, the user is assisted for all the data mining stages of the configured processes.
However, these approaches does not offer many ways to refine these processes, e.g., the possibility of adding or removing elements from the input data to be analysed is not supported.

As an example, Zorrilla et al.~\cite{Zorrilla2013a} developed, following the service-oriented methodology they propose, a high-level system for analysing student performance.
Using this system, teachers can, among other things, try to determine reasons why students fail, taking into account any available data of these students, e.g. activity in the e-learning platform, demographic information, or results in partial tests.
Nevertheless, the developed analysis system does not allow teachers to decide which data is used to analyse the students' performance, i.e., teachers cannot focus only on activity data to discover performance patterns related with how students use the platform.
If we wanted to incorporate this analysis, a data scientist would be again required to modify the tool.
Therefore, refinement support is somehow limited in this category.

\paragraph{EQ1.2: How is the decision maker assisted during each stage?}

\begin{table}
  \centering
  \caption{Assistance offered by each category during the analysis process.}
  \begin{tabular}{m{0.2\linewidth}m{0.7\linewidth}}
  	\toprule
  	Category                        & EQ1.2: Decision makers are assisted through ...                                                                        \\ \midrule
  	Workflows                       & building blocks that allow to graphically configure a complete analysis workflow.                               \\
  	SSBI                            & a user-friendly interface where high-level menus and wizards allow performing data-related tasks.                      \\
  	Black-box Components & tools that perform a concrete mining stage automatically.                                        \\
    Development Methodologies       & rules and frameworks to instantiate systems that can be used by non-expert users. \\ \bottomrule
  \end{tabular}
  \label{table:apaWhatAssistance}
\end{table}

We consider now how each category assists end users for the data mining stages it covers.

\emph{Workflow-based applications} provide building blocks that allow specifying data mining processes graphically.
Nevertheless, users are still responsible for selecting the right blocks for each analysis, and from their appropriate configuration and interconnection.
To accomplish these tasks, sound knowledge on data mining concepts is required.
Therefore, it can be concluded that workflow-based applications do not provide much assistance for the non-expert users we are focusing on this work.

\emph{SSBI tools} offer some wizards and interfaces to define data analysis tasks.
These tasks are mostly oriented to data reporting and descriptive analysis, although some tools have incorporated support for new data-related tasks over the last years, such as the \emph{Prediction Workbooks} offered by Watson Analytics\footnote{\url{http://bit.ly/predworkbooks}}.
As it happened with the workflow applications, users are still responsible for executing some low-level tasks, for which specialised knowledge might be required.
For instance, in the general case, users might need to aggregate or normalize data before they can be used as input for a data analysis task.

\emph{Black-box components} provide simple commands and interfaces that hide all low-level details of certain analysis tasks to the end user.
Therefore, these elements can be perfectly executed by users with no expertise in data mining techniques.
Nevertheless, only the data mining or data preprocessing stages, as already commented, are currently covered by these approaches.

Lastly, \emph{development frameworks} do not provide any support for executing data mining processes without the intervention of an expert.
These methodologies provide some development rules, often associated to a prebuilt infrastructure, to help create data analysis systems ensuring that the resulting products can be employed by end users without expertise in data mining tasks.
Therefore, a data mining expert is initially required to follow these rules and to configure appropriately the associated infrastructure, when it is provided.
Then, once the system is ready, end users can select data, execute different analysis tasks by themselves, and obtain results that can be easily interpreted.
Although this initial intervention of a data mining expert is always required, some of these approaches~\cite{santos_data_2013,espinosa_enabling_2015} aim to reduce this intervention as much as possible.

On the other hand, this initial intervention required in the development frameworks' approaches allows for the obtention of a system with higher accuracy than the achieved, for instance, when using black-box components, because this system has been adjusted to the particularities of the specific domain it is being deployed to.
This means that, assuming an extra cost, system accuracy can be increased.
This and other trade-offs are commented in the next section.


\subsubsection{Analysis of Trade-Offs}

This second stage of the evaluation procedure (see Section~\ref{sec:evaluationProcedure} and Table~\ref{table:tradeoffs}) aims to know how each one of the categories deals with the different trade-offs that are inherent to data mining processes.
Table~\ref{table:tradeoffsresults} summarises the results of this analysis.
A dash (``-'') appears as answer for those questions where an answer to a previous question invalidates them, e.g., EQ2.1 and EQ2.2.

\begin{sidewaystable}
  \centering
  \caption{Results of the quality attributes by category.}
  \begin{tabular}{lccm{2.5cm}m{3cm}}
  	\toprule
  	Quality Attribute                                     & Workflows &    SSBI     & Black-Box Components & Development Frameworks \\ \midrule
  	Adoption Cost                                         &           &             &                      &                           \\
  	EQ2.1. Can it be deployed without adaptations?        &    Yes    &     Yes     & Yes                  & No                        \\
  	EQ2.2. Can non-experts perform the adaptations?       &     -     &      -      & -                    & No                        \\ \midrule
  	Accuracy                                              &           &             &                      &                           \\
  	EQ2.3. Can the analysis reach expert-level accuracy?  &    Yes    &     No      & No                   & Yes                       \\
  	EQ2.4. Can the approach be tuned to improve accuracy? &    Yes    &     Yes     & No                   & No                        \\
  	EQ2.5. Can non-experts perform the tunings?           &    No     &     No      & -                    & -                         \\ \midrule
  	Completeness                                          &           &             &                      &                           \\
  	EQ2.6. What analysis techniques are available?        &    All    & Descriptive & Predictive           & All                       \\ \midrule
  	Evolvability                                          &           &             &                      &                           \\
  	EQ2.7. How easy it is to extend the approach?         &   Easy    &    Med.     & Hard                 & Med.                      \\
  	EQ2.8. Can new analysis techniques be included?       &    Yes    &     Yes     & No                   & Yes                       \\ \bottomrule
  \end{tabular}
  \label{table:tradeoffsresults}
\end{sidewaystable}

\emph{Workflow-based tools} provide building blocks that are designed to work with any input dataset, so they are can be used in any domain without prior adaptation.
However, as commented in the previous section, these building blocks are not designed to be employed by non-expert users, so a data scientist is always required to design the workflows that will implement a specific analysis process.

Building blocks come with a default configuration, so that they can be used with a minimum effort.
Nevertheless, this default configuration might not perform well for all domains~\cite{wolpert96}, provoking that the results of these defaults might not be as accurate as possible.
Defaults can be modified to increase the accuracy of the results, but this task requires a deep knowledge of data mining techniques, making necessary again the help of a data scientist.

Regarding evolution, new business questions might be addressed by means of creating new workflows, for which a data scientist would be required.
Similarly, existing workflows might be adapted to new requirements by changing the configuration and interconnection of their building blocks.
For instance, if after performing an analysis over a dataset, we wanted to repeat that analysis but just for a subset of the input data, this change might be addressed by adding a filter block after the data loading step.
Also, new data mining techniques could be incorporated to these tools using extension mechanisms that support the addition of new building blocks, which wrap custom data mining tasks.
As before, the intervention of an expert would be required perform any of these tasks.

\emph{SSBI tools} are available as generic solutions that, in most cases, can be deployed in any context as are and used by decision makers without advanced expertise in data analysis.
Using these tools, these decision makers would define the required analysis process by themselves.
Therefore, these tools can be classified as domain-independent solutions that can be set up with a relative low effort, and without the intervention of a data scientist.
Nevertheless, as previously described (see Section~\ref{section:background}), most SSBI tools, at the time of writing this work, focus mainly on offering reporting capabilities, including basic descriptive analytics processes that can be incorporated to these reports.

The accuracy of SSBI tools might be compromised by the powerfulness of the offered functionality.
These tools try to be understandable by a lot of different users with heterogeneous expertise, so some advanced functionalities might be not present to favour the simplicity and amenability of the tool.
As an example, for a data preprocessing stage, users of a SSBI tool can only choose between the available list of cleaning tasks, which may not be as large as the plethora of libraries and specialised tools than an expert might employ for this purpose.
While, to some extent, this issue might also apply to workflow applications, in our review we have observed that workflow apps are way more complete in terms of offered functionality than SSBI solutions.

In terms of evolvability, the analysis processes initially designed by decision makers can also be modified and refined by themselves.
In addition, if new data mining techniques were required to answer new business questions, these techniques might be incorporated into some of these tools through custom scripts.
For instance, if we wanted to incorporate support for calculating association rules in one of these tools, we should write a script that implements or invokes the corresponding algorithm.
Unfortunately, this kind of task demands the intervention of data mining experts.

\emph{Black-box components} are devised to work as are for any input, and with no prior adaptation work required.
However, as commented in the previous section, these approaches do no cover all stages of a data mining process.
For instance, most approaches focus exclusively on the mining stage, and specifically in offering prediction analysis tools.
This implies that, for instance, end users are in charge of selecting, cleaning and formatting the data of interest for an analysis into the format accepted by the employed black-box prediction tool. 
These tasks will be often too far away from the capabilities of decision makers.
Therefore, although these approaches can be easily included in new contexts to cover some stages of the analysis process, the intervention of data scientists might be required in order to perform the remaining stages.

This black-box approaches do not take into account any specificities of the application domain during the analysis, which might decrease the accuracy of the obtained results.
This lack of adaptations to each domain might return worse results than the ones an expert may achieve.
In addition, the black-box nature of these tools might make unfeasible to adapt these applications to a specific domain, even with expert intervention.
For instance, most of the tools provide a way to create prediction models, but we might not be able to slightly modify how the prediction model is built, e.g., by deciding which features are more relevant for the prediction.
Consequently, these applications cannot be extended to support new user needs, and new analysis techniques cannot be easily incorporated.

\emph{Development frameworks} are designed to be used in any context, so they can be classified as domain-independent initially.
Nevertheless, these approaches often provide an infrastructure that needs to be modified to fit in with the particularities of a specific domain. 
For instance, in Santos et al.~\cite{santos_data_2013}, some metadata needs to be specified by a data mining expert to indicate how the different elements of a domain ontology should be processed. 
Therefore, these approaches need  some previous work before being deployed in a new context, and this work must be carried out by an expert. 

The counterpart of the higher cost of these approaches is that the initial intervention of experts allows taking into account any relevant details of the application domain, which contributes to increase the degree of accuracy of these solutions to something very similar to what an expert might achieve.
For instance, in the educational example of Zorrilla and Garc\'{i}a-Saiz~\cite{Zorrilla2013a}, although teachers use wrapped processes that are not configurable, these processes have been adapted to the educational domain, which may improve the quality of the results when compared with, for instance, commands of black-box components.

These tools can also be extended to support new business needs.
For instance, in Santos et al.~\cite{santos_data_2013}, the domain ontology might be used to select different subsets of data as we are gaining insights in a domain, and we want to find answers to more precise questions.
Nevertheless, not all these approaches support this kind of refinement, and the intervention of experts might be required depending on what new user needs we want to address.
Similarly, new data mining techniques can be incorporated to these methodologies and their associated infrastructure with the help of an expert.

With the previous paragraph we finished the description of the obtained results in the evaluation.
Next section answers the research questions that we formulated as the objectives of this work.

\section{Answers to research questions}
\label{sec:answers}

During the evaluation procedure, we gathered enough evidence to provide answers for our research questions (see Section~\ref{sec:researchquestions}).
These answers can be found below.

\paragraph{RQ0. What approaches tackle the problem of data mining democratisation?}

We identified four different categories of approaches that collaborate in the data mining democratisation field: \emph{workflow applications}, \emph{Self-Service Business Intelligence solutions}, \emph{black-box components} and \emph{development frameworks}.
More details of these categories and the approaches belonging to them can be found in Section~\ref{sec:resultssummary}.
Although there seems to be an interest in this field, the amount of articles found in the academia is still small, as not many approaches address this subject yet.
On the other hand, important enterprises, such as IBM or Microsoft, are starting to perform a non-negligible effort to provide decision makers with user-friendly solutions for performing data exploration and analysis.
This means that data mining democratisation is being considered an important issue, which will need to be addressed more in-depth in the near future.

\subsection{RQ1. When using the approaches identified in the previous question, what actions do decision makers need to carry out to analyse a dataset?}

These actions vary highly depending on the kind of approach selected.
In workflow-based applications, users connect and configure pre-built blocks to specify a data mining process.
In SSBI solutions, users navigate through a set of high-level user interfaces and wizards to configure an analysis process, more based on visualization and reporting than in data mining.
Black-box components do not provide support for all stages of the analysis process, so the user needs to carry some of them without any assistance.
For instance, user often needs to retrieve, format and clean the data to analyse.
For those stages that are automated, users just need to invoke some commands that hide the low-levels details of their execution.
In development frameworks, an initial minimum intervention of a data-scientist is required.
After that, users can operate the systems by themselves through interfaces abstracted from low-level details.

\subsection{RQ2. What technical knowledge is required to carry out the actions?}

Workflow-based applications require a sound knowledge of data mining techniques, since the offered building blocks are more oriented to data scientists rather than for non-expert users.
On the other hand, SSBI solutions are mostly designed to be used by non-experts, so no technical knowledge is initially required.
Nevertheless, if advanced analysis are wanted, i.e., going beyond basic reporting, some technical knowledge might be necessary.
Black-box components do not require any technical knowledge for the stages of the data mining process they address.
Development frameworks require an initial configuration of a certain infrastructure.
This initial configuration has to be carried out by an expert but, once it is completed, the resulting system can be operated by non-expert users.

\subsection{RQ3. Can non-expert users make use of data mining tools and techniques by themselves?}

In the light of the answers to previous questions, the answer is no.
Workflow-based applications are not designed to be employed by non-experts.
SSBI solutions offer a limited support for advanced data mining techniques.
When these techniques are addressed, SSBI solutions often face the same problems as workflow-based applications, i.e., they do not provide suitable solutions for non-expert users.
Black-box components can be used by non-experts, but they do not address the whole data mining process.
Finally, development frameworks require the initial intervention of a data scientist, although they aim to reduce this intervention to the minimum.

\subsection{RQ4. What trade-offs need to be considered for achieving data mining democratisation?}

\emph{Genericity versus accuracy} and \emph{accuracy versus cost} seem to be prominent trade-offs for the studied approaches.
Those approaches that are absolutely domain-independent, such as the black-box components, do not require the intervention of experts, so they can be adopted with a relatively low-cost.
Nevertheless, since they cannot be optimised for taking into account the particularities of any domain, their results might be not as accurate as possible.
In some cases, accuracy might be seriously compromised and the system would not fulfil end-user needs.
On the other hand, the need of experts to include these domain-specific customisations implies an increase in adoption cost.

\subsection{RQ5. What should be improved in current state-of-the-art so that decision makers can properly analyse datasets by themselves?}

\emph{Black-box components} seem to be the more promising attempt to data mining democratisation, since the intervention of experts, for those stages these approaches address, is not required.
Nevertheless, as previously commented, accuracy is sometimes compromised.
Therefore, it would be desirable to find techniques that help to automatically select and configure the algorithms that best fit in with the particularities of each domain.
Some works, such as Reif et al.~\cite{reif_automatic_2014} or Billali et al.~\cite{Bilalli2017}, offer some initial results in this direction.
These works belong to the area of \emph{metalearning}~\cite{Brazdil2009}, where meta-prediction models are built to help select the best algorithms for an analysis.
Other possible solutions are \emph{autoconfigurable} or \emph{parameter-less} techniques~\cite{Zorrilla2011,Feurer2015}, which try to tune their parameters automatically to offer the best possible results.
Nevertheless, more research work is needed in this area to allow \emph{black-box components} to be completely automated while avoiding any noticeable accuracy loss, facilitating data mining democratisation.

Metalearning and autoconfigurable algorithms have provided interesting solutions for some specific problems and areas, such as classification and regression problems, but there is still a lack of generic solutions that can be universally applied.
While we wait for these global solutions, the possibility of tuning, with minimum expert intervention, an initially generic application, such as development frameworks do, seems to be an interesting and pragmatic solution to the \emph{accuracy versus cost} trade-off.
Therefore, the design of generic analysis frameworks that can be instantiated in a concrete domain, with a minimum expert intervention, should be studied more in-depth.

Most of the analysed work focus on data mining algorithm selection and execution, e.g.~\cite{ankerst_towards_2000,reif_automatic_2014}.
Only two primary studies, belonging to the black-box components category, dealt with issues on the preprocessing stage~\cite{kanter2015,Bilalli2017}.
This shows a lack of work addressing the \emph{data obtention} and \emph{preprocessing} stages of the data mining process, even when these stages have been demonstrated critical for the outcome of an analysis~\cite{crone2006,munson2012}.
Therefore, more research work would be needed in these stages.

 
\section{Conclusions}
\label{section:conclusions}

This article has presented a systematic review that aims to identify how far we are from an effective democratisation of data mining.
To achieve this goal, we followed the review protocol proposed by Kitchenham and Carters~\cite{Kitchenham2007}, which we complemented with the \emph{snowballing} technique as proposed by Jalali and Wohlin~\cite{Jalali2012}. 
In this review, we considered as primary studies both research work and state-of-the-art tools.
This combined analysis gave us the complete picture of the field, including what is available now and what new techniques should be expected to be adopted in the upcoming years.
During the review, 559 research papers and 138 tools were initially considered, from which we selected 15 articles and 28 tools as primary studies, adding up to 43 (see Table~\ref{table:categories}).

In terms of quantity, it seems that there is a considerable interest in the industry in offering solutions for data mining democratisation.
With respect to the academia, although there are some publications in this field, most of them refer to solutions for very specific problems, and as such they are difficult to generalize. 
The number of solutions found that actually try to improve the situation of data mining democratisation from a general perspective was not that large.

The selected primary studies were grouped in four different categories: (1) \emph{Workflow-based applications}; (2) \emph{Self-Service Business Intelligence (SSBI) solutions}; (3) \emph{Black-box components}; and (4) \emph{Development frameworks}.
Each category was then analysed against a well-defined systematic evaluation protocol.

The evaluation concluded that \emph{workflow-based applications} are not designed to be employed by non-expert users, whereas \emph{SSBI solutions} offer limited support for advanced data mining techniques at the time of writing this review.
\emph{Black-box components} are perfectly usable by non-experts, but they only address some stages of the data mining process, and might exhibit accuracy flaws.
\emph{Development frameworks} try to solve this accuracy problem by means a controlled and reduced intervention of a data mining expert, who would perform domain-specific customisations for a particular setting.
Moreover, it can also be concluded that little attention has been paid to the \emph{data acquisition} and \emph{preprocessing} stages.
To sum up, it can be stated that, although there are some promising initial steps, we are still far from data mining democratisation.
 
\section*{Acknowledgements}

This work has been partially funded by the Government of Cantabria (Spain) under the doctoral studentship program from the University of Cantabria, and by the Spanish Government under grant TIN2017-86520-C3-3 R.

\appendix

\section{Comments on Ad-Hoc Applications}
\label{sec:domainspecificapps}
As previously commented in the selection strategy for scientific databases (see Section~\ref{sec:selectionCriteriaAcademia}), we excluded a special category of papers from our evaluation process.
This category is composed of \emph{ad-hoc applications}, which were developed to solve a concrete analysis problem on a specific domain.
Proof of this is that, for instance, we found several applications focusing on the analysis of large, multidimensional datasets~\cite{Klenk2009,Omta2016,Proietti2016}; on applying Inductive Logic Programming~\cite{Luu2012, Santos2011}; or on analysing time-series data~\cite{Chittaro2003,Guyet2007,Mellis2017}.
However, each application was developed independently, and adapted to a concrete domain (e.g medicine for~\cite{Klenk2009}, biomedicine for~\cite{Omta2016}, and biomolecular for~\cite{Proietti2016}).
This kind of conventional, personalised development is the one that escaped from the focus of this review.

Nevertheless, we decided to include a short summary of these applications, because they made a special effort to be usable by domain experts without technical knowledge in data mining techniques.
We considered that this effort could be transferred to other applications, so they are of interest for the ultimate goal of democratising data mining.

\begin{sidewaystable}
  \centering
  \caption{Characteristics of encountered ad-hoc applications for non-experts.}
  \begin{tabular}{llm{14cm}}
  	\toprule
  	Reference                & Domain       & Contribution to data mining democratisation                            \\ \midrule
  	\citet{Chittaro2003}     & Medicine     & A circular control panel for the management of 2D and 3D visualisations of time series data.    \\
  	\citet{Garcia2011}       & Education    & An association rule learning tool for the analysis of educational data that allows sharing the obtained rules between courses of similar nature. \\
  	\citet{Guyet2007}        & Medicine     & A tool to perform pattern matching visually in time series data (e.g. respiratory data). \\
  	\citet{Jugo2015}         & Education    & Web-based interface for clustering analysis and for guiding users in a learning path.        \\
  	\citet{Kamdar2014}       & Biomedicine  & A high-level system to formulate data queries over ontology-based data sources.                     \\
  	\citet{Kamsu-Foguem2012} & Medicine     & A machine learning-enhanced monitor that automatically determines the most relevant indicators to show in a constrained-size display (e.g. medical intensive care units).             \\
  	\citet{Klenk2009}        & Medicine     & A system to perform case-based reasoning (e.g. history-based survival analysis of a patient).               \\
  	\citet{Luu2012}          & Genomics     & An Inductive Logic Programming (ILP) system that analyses data and gives rule-based explanations understandable by non-experts in data mining. \\
  	\citet{Peng2015}         & Agriculture  & A geographic system that allows obtaining map-based visualisations of different indicators of interest (e.g. drought maps of certain areas)\\
  	\citet{Mellis2017}       & Electronics  & A high-level solution for the analysis of raw sensor data. \\
  	\citet{Omta2016}         & Biomedicine  & Web service for analysing High-Content (i.e. large and multidimensional) datasets. \\
  	\citet{Proietti2016}     & Biomolecular & Web service for the analysis and comparison of different multidimensional datasets.       \\ 
    \citet{Santos2011}       & Biomolecular & An ILP system for multi-relational data mining (similar objectives to~\citet{Luu2012}).  \\ \bottomrule
  \end{tabular}
  \label{table:domspecifapps}
\end{sidewaystable}

Applications of this category offer solutions for very concrete problems.
These applications target heterogeneous domains, including molecular biology, medicine, biomedicine, genomics, education, electronics and agriculture.
Table~\ref{table:domspecifapps} shows the domain of each application, and the specific help they provide for non-experts and that we consider relevant for the general purpose of data mining democratisation.

For instance, \citet{Chittaro2003} applies time-series analysis methods for the monitoring of haemodialysis processes.
This method is integrated in a user-friendly system, which was developed by data mining experts.
The system allows clinicians to envision the evolution over time of different metrics presented in 2D or 3D user-friendly visualizations.
The application includes a special circular control panel, which allows clinicians to tune the visualization and analysis parameters.
Despite being an application focused on the study of haemodialysis, part of this solution could be transferred to the visualization and analysis of time series data coming from other domains.

\bibliographystyle{elsarticle-num-names}

\end{document}